\def\diff{\mathrm{d}}

\def\A{\mathrm{A}}
\def\B{\mathrm{B}}
\def\C{\mathrm{C}}
\def\D{\mathrm{D}}

\def\vecr{\mathbf{r}}
\def\vecp{\mathbf{p}}
\def\Ntest{N_{\textrm{test}}}
\def\Icoll{I_{\textrm{coll}}}
\def\Epot{E_{\textrm{pot}}}
\def\Etot{E_{\textrm{tot}}}
\def\ECoulpot{E^{\textrm{Coul}}_{\textrm{pot}}}
\def\Esym{E_{\textrm{sym}}}
\def\Csurf{C_{\textrm{surf}}}
\def\Csym{C_{\textrm{sym}}}
\def\sigmaNN{\sigma_{\textrm{NN}}}

\newcommand{\beq}{\begin{equation}}
\newcommand{\eeq}{\end{equation}}
\def\bbsty#1#2#3{{\bf #1}, #2 (#3)}	

\documentclass{ws-rv961x669}
\usepackage{ws-rv-van}     
\usepackage{ws-rv-thm}     
\usepackage{subfigure}     
\makeindex

\begin{document}

\chapter[Mean-field instabilities and cluster formation in nuclear reactions]{\Large Mean-field instabilities and \\cluster formation in nuclear reactions}\label{ra_ch1}
\vspace{-50pt}

\author[M. Colonna, P. Napolitani, V. Baran]{M. Colonna$^{1}$, P. Napolitani$^{2}$, V. Baran$^{3}$}

\vspace{10pt}

\address{
$^1$INFN-Laboratori Nazionali del Sud, I-95123, Catania, Italy\\  
$^2$ Institut de Physique NuclÃ©aire, CNRS/IN2P3, Universit\'e Paris-Sud, Universit\'e Paris-Saclay, 91406 Orsay cedex, France\\
$^3$ Faculty of Physics, University of Bucharest, Romania\\ 
}

\begin{abstract}
\begin{center}
\line(1,0){330}
\end{center}
We review recent results on intermediate mass cluster production in
heavy ion collisions at Fermi energy and in spallation reactions.
Our studies are based on modern transport theories, employing effective
interactions for the nuclear mean-field and incorporating two-body
correlations and fluctuations.
Namely we will consider the Stochastic Mean Field (SMF) approach and
the recently developed Boltzmann-Langevin One Body (BLOB) model.
We focus on cluster production emerging from the possible occurrence of
low-density mean-field instabilities in heavy ion reactions.
Within such a framework, the respective role of one and two-body effects,
in the two models considered, will be carefully analysed.
We will discuss, in particular, fragment production in central and
semi-peripheral heavy ion collisions, which is the object of many
recent experimental investigations.  Moreover, in the context of spallation
reactions, we will show how thermal expansion may trigger the development
of mean-field instabilities, leading to a cluster formation process
which competes with important re-aggregation effects.
\begin{center}
\line(1,0){330}
\end{center}
\end{abstract}


\body

\section{Introduction}
Nuclei are quantum many-body systems consisting of protons and neutrons strongly 
bound together. Understanding the properties of such complex systems 
in terms of their constituent particles and the interaction among them
is a true challenge.    
The original quantal  many-body problem, is often approached
adopting the mean-field approximation, yielding a so-called 
effective interaction \cite{Neg,Dec,Rein,Bend}. %
This scheme is quite successful in describing the features of the
nuclear ground state, which exhibits a shell structure, where  
nucleons move almost independently in an averaged field (i.e. the mean field), analogously
to an atomic system. Other relevant aspects of the nuclear structure, such as the
excitation modes linked to single-particle motion   
and/or to the emergence of collective phenomena, are well understood within such a 
framework.
However, nuclear systems also manifest unique features, different from other
many-body systems. Indeed,    
a nucleus is
a self-bound system, where spatial correlations among nucleons can be 
rather strong. As a consequence,  assembling
and disassembling of nucleons may occur in several ways.  Moreover, owing to the
weak dependence of binding energy and saturation density on the mass number,
disassembling of nucleons and re-aggregation in new configurations may happen
at relatively low excitation energy.  
As  well known, these characteristics lead to the appearing 
of  cluster structures, in
which a nucleus is divided into several subunits (clusters) and nucleons are confined
within each cluster.  This already happens in the ground state configuration 
of selected nuclei \cite{Ebran2014,Kanada2012}.     
However, this fragmentation process becomes quite important in excited systems, 
as those formed in intermediate-energy heavy-ion collisions, where several 
clusters and nuclear fragments are produced from a hot source whose
excitation energy is typically comparable to the binding energy (per nucleon)
of a nucleus.
Nuclear clusters are predicted to appear also in equilibrated nuclear matter below saturation density, i.e. in conditions encountered in the inner crust of neutron stars
and/or along supernova explosion processes \cite{Lattimer2000}.   

Suitable extensions of mean-field models have to be introduced to take explicitly
into account the effects of relevant interparticle correlations.
Focusing on nuclear dynamics, 
an intense theoretical work 
on correlations and density fluctuations has started in the past years, 
also stimulated by the availability of 
large amounts of experimental data on fragment formation in intermediate 
energy heavy ion collisions and the possibility to observe 
volume (spinodal) instabilities,  
thus assimilating the nuclear disassembling to
the occurrence of liquid-gas 
phase transitions \cite{Frankland,Moretto,Dagostino,EPJA,rep,Ono}.
 
The dynamics of nuclear collisions at intermediate
energy is often investigated within the framework of semi-classical transport theories, such as the Nordheim approach, in which the Vlasov equation for the
one-body phase space density, $f({\bf r}, {\bf p},t)$, is supplemented with a Pauli-blocked
Boltzmann collision term \cite{Bertsch_rep,Aldo_rep},
which accounts for the average effect of the two-body residual interaction 
(quantal correlations are not included).  
The basic ingredients that enter the resulting transport 
equation, often called  Boltzmann-Uehling-Uhlenbeck (BUU) 
or Boltzmann-Nordheim-Vlasov (BNV)
equation,  
are
the self-consistent mean-field potential and the two-body scattering cross sections.
In order to introduce fluctuations and further (many-body) correlations in transport theories, 
a number of different avenues have been taken, that can
be essentially reconducted to two different classes of models. 
One is the class of molecular dynamics (MD) models \cite{Ono,md,feld,ONOab,Pap,FFMD} while the other
kind is represented by stochastic mean-field approaches 
\cite{Ayik,Ayik1,Randrup,rep}.
In molecular dynamics models the many-body state is represented 
by a simple product wave function, with or without antisymmetrization.
The single particle wave functions are assumed to have a fixed Gaussian shape.
In this way, though nucleon wave functions are supposed to be
independent (mean-field approximation), the use of localised wave packets
induces many-body correlations both in mean-field propagation 
and hard two body scattering (collision integral), which 
is treated stochastically.
Hence this way to introduce many-body correlations 
and produce a trajectory branching, leading to a variety of clustered 
configurations,
 is essentially based on the use of empirical Gaussian
wave packets. 

This localisation of the nucleon wave packet has shown to be quite successful in
describing the clustered structures characterising the ground state 
of several light nuclei \cite{Kanada2012}.    
However, as far as nuclear dynamics is concerned, 
while the wave function localisation appears appropriate
 to describe final fragmentation channels,  
where each single particle wave function should be localised within a
fragment,  
the use of fixed shape localised wave packets in the full dynamics 
could affect the correct description of  
one-body effects, such as spinodal instabilities and zero sound 
propagation \cite{FFMD,Akira_comparison}. 

On the other side, in the so-called stochastic mean-field approaches, 
which we will adopt in the following, 
the stochastic extension of the transport treatment for the one-particle
density is obtained by introducing a stochastic term representing the 
fluctuating part of the collision integral \cite{Ayik,Ayik1,Randrup}, in close analogy
with the Langevin equation for a Brownian motion.
This can be derived as the next-order correction, in the equation describing
the time evolution of $f$, with respect to the standard 
average collision integral, leading to the Boltzmann-Langevin (BL) equation. 
Thus, the system is still described solely in terms of the reduced one-body density
$f$, but this function experiences a stochastic time evolution in response
to the random effect of the fluctuating collision term.
In this way density fluctuations, resulting from many-body correlations, 
 are introduced, that are amplified when 
instabilities or bifurcations occur in the dynamics.  
This procedure is suitable for addressing multifragmentation phenomena, where clusters 
emerge from the growth of density inhomogeneities, driven by the unstable mean-field. 
However, the fluctuations introduced in this way are not strong enough to
fully account for the production of 
light clusters, which are loosely bound by the mean-field and would require
a strong localisation of the nucleon wave packet (as in MD approaches).
In the following we
will concentrate on the production mechanism of medium size ($Z>2$) clusters,
emerging in nuclear reactions from the disassemby of excited (warm and/or
diluted) sources.  In particular, we will illustrate how the clustering process
is ruled by the interplay between the presence of many-body correlations
and the role of mean-field instabilities, which amplify the cluster 
aggregation process in low-density matter.
Moreover, we will discuss the sensitivity of the cluster features
to relevant properties of the nuclear effective interaction.  
The paper is organised as follows: 
in Section 2 we review the theoretical framework. Section 3 is devoted to a
survey of results for central heavy ion reactions at Fermi energies. 
In Section 4 we discuss semi-peripheral reactions and neck dynamics, whereas
Section 5 presents some results for fragmentation in spallation reactions. 
Finally conclusions and perspectives are drawn in Section 6. 

\section{Theoretical description of nuclear reactions}

\subsection{The Stochastic Mean Field (SMF) model}


Nuclear reactions can be modelled by solving transport equations
based on 
mean field theories, with short range (2p-2h) correlations included via hard nucleon-nucleon
elastic collisions and via stochastic forces, selfconsistently
evaluated from the mean phase-space trajectory. 

The SMF model \cite{Salvo} can be considered as an approximate tool to solve 
the  
BL equation \cite{Ayik}:
\begin{equation}
{{df}\over{dt}} = {{\partial f}\over{\partial t}} + \{f,H\} = \Icoll[f] 
+ \delta I[f],
\label{eq1}
\end{equation}
 where $f({\bf r},{\bf p},t)$ is the one-body distribution function, 
the semi-classical analog of the Wigner transform of the one-body density matrix, 
$H({\bf r},{\bf p},t)$ the mean field Hamiltonian, 
$\Icoll$ the two-body collision term 
incorporating the Fermi statistics of the particles,
and 
$\delta I[f]$ its fluctuating part. 
The coordinates of isospin are not shown for brevity.
Eq.(1) is solved adopting the test particle method, i.e. each nucleon is
associated with a given number, $\Ntest$, of test particles.

It should be noticed that in the BUU/BNV models, the fluctuating term $\delta I[f]$
is neglected \cite{Bertsch_rep,Aldo_rep}.  
In the present SMF treatment we project the fluctuations of the distribution
function, generated by the stochastic collision integral in Eq.(1),  
on the coordinate space and consider local density fluctuations, 
which could be implemented
as such in a numerical calculation. 
We make the further
assumption of local thermal
equilibrium, thus being able to derive analytic 
expressions for the density fluctuations \cite{Salvo}.

When instabilities are encountered along the reaction path, 
the evolution of the fluctuation ``seeds'' introduced by the SMF method is then
determined by the dissipative dynamics of the BNV evolution, 
allowing 
the system to choose its trajectory through 
the fragmentation configuration. In this way we create a series of
``events'' in a nuclear collision, which can then be analysed and sampled 
in various ways.


\subsection{Boltzmann-Langevin dynamics: the BLOB model}

The implementation of the full structure in phase space of the original BL term
can still be considered as an important goal.
In fact, this allows one to treat a more general class of phenomena, where 
the correct description of fluctuations and correlations in  ${\bf p}$ space
is essential (such as fragment velocity correlations for instance).
We also stress the general interest of this effort. Indeed transport phenomena occur in many physical
systems, for which a more precise description of the time evolution of the one-body distribution
function, including the effect of many-body correlations, 
would be important.

The recently introduced BLOB~\cite{Napolitani2013} model
aims at solving the full BL equation in phase space. 
While the BLOB model inherits the mean-field description from the SMF model, a different approach in treating the collision integral $\bar{I}[f]$ and the fluctuation term $\delta I[f]$ (i.e. the right hand side of eq.~\ref{eq1}) is employed.
As it will be discussed in the following, the numerical implementation of the BLOB approach imposes that the residual term $\bar{I}[f]+{\delta I[f]}$ agitates extended portions of the phase space in each single scattering event.
However, the model differs substantially from an earlier similar strategy, used in the Bauer-and-Bertsch approach~\cite{Bauer1987}, because it constrains the fluctuating term ${\delta I[f]}$ to act on phase-space volumes with the correct occupation variance.
Such constraint avoids that the Pauli blocking could be violated, and it imposes to pay special attention to the metrics of the phase space (see discussion in ref.~\cite{Chapelle1992}).
Nuclear-matter calculations in a periodic box, in one dimension~\cite{Rizzo2008} and in three dimensions~\cite{Napolitani_IWM2014} have shown that the BLOB approach describes, for low-density nuclear matter,  the growth rate of unstable modes in correct connection with the form of the mean-field potential, as ruled by the dispersion relation~\cite{Colonna1994}. 
	The BLOB model for heavy-ion collision is constructed as based on this efficient description of the dispersion relation. 


	The solution of the BL equation in full phase space is obtained by replacing the conventional Uehling-Uhlenbeck average collision integral by a similar form where one binary collision does not act on two test particles $a,b$ but it rather involves extended phase-space agglomerates of test particles of equal isospin A$={a_1,a_2,\dots}$, B$={b_1,b_2,\dots}$ to simulate wave packets:
\begin{equation}
	{\bar{I}[f]}+{\delta I[f]}
	= g\int\frac{\diff\vecp_b}{h^3}\,
	\int \diff\Omega\;\;
	W({\scriptstyle\A\B\leftrightarrow\C\D})\;
	F({\scriptstyle\A\B\rightarrow\C\D})\;,
\label{eq2}
\end{equation}
where $W$ is the transition rate, in terms of relative velocity between the two colliding agglomerates and differential nucleon-nucleon cross section
\begin{equation}
	W({\scriptstyle\A\B\leftrightarrow\C\D}) = |v_\A\!-\!v_\B| \frac{\diff\sigma}{\diff\Omega}\;,
\label{eq3}
\end{equation}
and $F$ contains the products of occupancies and vacancies of initial and final states calculated for the test-particle agglomerates
\begin{equation}
	F({\scriptstyle\A\B\rightarrow\C\D}) =
	\Big[(1\!\!-\!\!{f}_\A)(1\!\!-\!\!{f}_\B) f_\C f_\D - f_\A f_\B (1\!\!-\!\!{f}_\C)(1\!\!-\!\!{f}_\D)\Big]\;.
\label{eq4}
\end{equation}
	At each interval of time, by scanning all phase space in search of collisions, and by redefining all test-particle agglomerates accordingly in phase-space cells of volume $h^3$, nucleon-nucleon correlations are introduced.
	Since $\Ntest$ test particles are involved in one collision, and since those test particles could be sorted again in new agglomerates to attempt new collisions in the same interval of time as far as the collision is not successful, the nucleon-nucleon cross section contained in the transition rate $W$ should be divided by $\Ntest$: $\sigma = \sigmaNN / \Ntest$.
	Special attention should be payed to the metrics when defining the test-particle agglomeration: the agglomerates are searched requiring that they are the most compact configuration in the phase space metrics which does neither violate Pauli blocking in the initial and in the final states, nor energy conservation in the scattering.
	The localisation in momentum space makes the collisions more effective in agitating the phase space, and the localisation in coordinate space is needed to keep hydrodynamic effects like the flow dynamics. 

	The correlations produced through this approach are then exploited within a stochastic procedure, which consists in confronting the effective collision probability $W\times F$ with a random number.
	As a consequence, fluctuations develop spontaneously in the phase-space cells  of volume $h^3$ with the correct fluctuation amplitude.
	A precise shape-modulation technique~\cite{Napolitani2012} is applied to ensure that the occupancy distribution does not exceed unity in any phase-space location in the final states. 	This leads to a correct Fermi statistics for the distribution function $f$, in terms of mean value and variance.

\subsection{Ingredients of the calculations}

Within our framework, the total energy of the system
can be written as:
$$ \Etot = {{1} \over {\Ntest}}\sum_i p_i^2/(2m) + \int d\vecr~\rho(\vecr) \Epot(\rho_n,\rho_p) $$ 
\begin{equation}
 + \int d\vecr~\rho_p(\vecr) \ECoulpot(\rho_p)/2 ,
\end{equation}
where
$\rho_n , \rho_p$ denote neutron and proton densities, $\rho = \rho_n + \rho_p$, $\Epot$ is the potential
energy per nucleon, connected to the mean-field interaction, and  $\ECoulpot$
denotes the Coulomb potential.

Effective interactions, associated with a given Equation of State (EOS) 
can be considered as an input of all transport codes.
We adopt a soft isoscalar EOS (compressibility $K = 200$ MeV).
We notice that the considered compressibility value is favoured e.g. 
from flow, monopole oscillation and multifragmentation studies \cite{rep,Borderie}. 
The choice considered corresponds
to a Skyrme-like effective interaction, namely $SKM^*$, for which we take the effective
mass as being equal to the nucleon bare mass. 
Then $\Epot$ can be written as: 
\begin{equation}
\Epot(\rho)  = {A \over 2} {\tilde \rho} +  {B \over {\sigma+1}} 
{\tilde \rho}^\sigma + {\Csurf \over {2\rho}} (\nabla\rho)^2
+ {1 \over 2} \Csym(\rho){\tilde \rho} \beta^2,
\end{equation}
where ${\tilde \rho} = \rho / \rho_0$ ($\rho_0$ denotes the saturation density), $A = -356~$MeV,
$B = 303~$MeV, $\sigma = 7/6$.  
We notice that surface effects are automatically introduced 
in the dynamics
when considering finite width wave packets
for the test particles employed in the numerical resolution.
An explicit surface term is also added (third term of Eq.(6)) and 
tuned in such a way that the total surface energy reproduces 
the surface energy of nuclei in the ground state
\cite{TWINGO}.  This procedure yields $\Csurf = -6/\rho_0^{5/3}~$MeV~fm$^5$.
The last term of Eq.(6) represents the potential part of the symmetry energy
per nucleon, with $\beta = (\rho_n - \rho_p) / \rho$. 

For some of the reaction mechanisms analyzed with SMF and BLOB,
the sensitivity of the simulation results, and of the corresponding
cluster features, is tested against 
different choices of the density dependence of $\Csym$:
the asysoft,  $\displaystyle C(\rho) = {\rho_0} (482-1638 \rho) $MeV,
the asystiff, $\displaystyle C(\rho) = 32 $MeV,
and the asysuperstiff, $\displaystyle C(\rho) =32 \frac{2 \rho}{\rho+\rho_0} $MeV
\cite{epja}.
The value of the symmetry energy,
$\displaystyle \Esym/A = {\epsilon_F \over 3}+{1 \over 2} \Csym(\rho){\tilde \rho}$,
at saturation, as well as
the slope parameter, $\displaystyle L = 3 \rho_0 \frac{d \Esym/A}{d \rho} |_{\rho=\rho_0}$,
are reported in Table \ref{table1} for each of these asy-EOS
($\epsilon_F$ denotes the Fermi energy).
Just below the saturation density
the asysoft parameterisation exhibits a weak variation with density, while the asysuperstiff shows
a rapid decrease. 
\begin{table}
\begin{center}
\begin{tabular}{|l|r|r|r|r|r|} \hline
asy-EoS       & $\Esym/A$    & L(MeV)  \\ \hline
asysoft       &     30.                  & 14.   \\ \hline
asystiff      &     28.                  & 73.   \\ \hline
asysupstiff   &     28.                  & 97.   \\ \hline
\end{tabular}
\caption{The symmetry energy at saturation (MeV) and the slope parameter (MeV)
for the three asy-EOS considered (see text).} 
\label{table1}
\end{center}
\end{table}
Momentum-dependent effective interactions may also be implemented into Eq.(1) \cite{baranPR,Joseph}.


Two-body correlations are taken into account through the collision integral,
r.h.s of  Eq.(1). 
The main ingredient entering this process is the nucleon-nucleon cross section $\sigmaNN$, for which
we use the free isospin, angle and energy dependent values
or in-medium modified parameterisations~\cite{DanielewiczCoupland}.

It should be noticed that the procedure adopted to solve the collision integral, which employs random
numbers, is stochastic.  In SMF, owing to the fact that collisions are treated
for pairs of test particles, fluctuations are reduced by $1 / \Ntest$.
Thus an explicit fluctuation term is needed, as indicated in Eq.(1) and explained
above, to account for the stochastic nature of the nucleon-nucleon collision process. On the other hand, in BLOB collisions are treated directly for nucleons, 
i.e. for test particle agglomerates,
independently of the value of $\Ntest$. 

\section{Survey of results for central collisions}
%
%
%

\begin{figure}[b!]
\centerline{
  \subfigure[]
     {\includegraphics[angle=0, width=.58\textwidth]{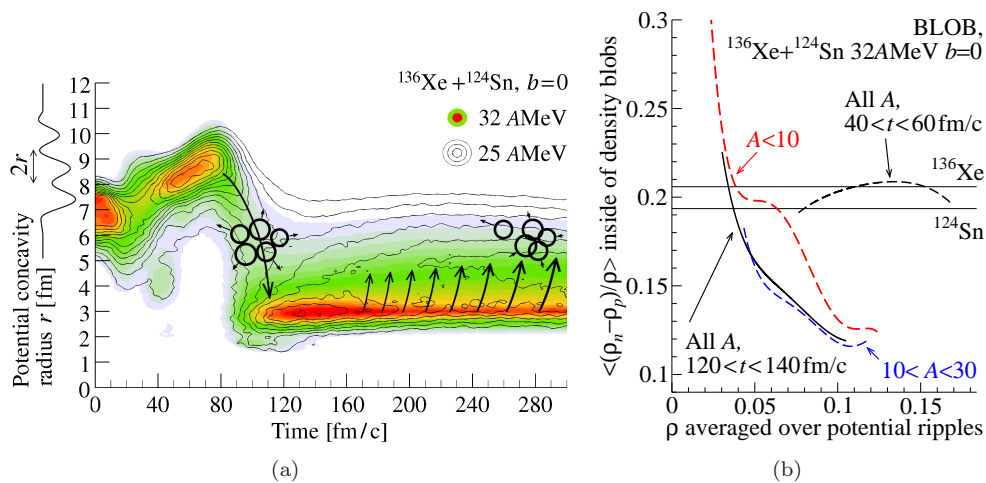}\label{fig_phase_trans_a}}
  \hspace*{4pt}
  \subfigure[]
     {\includegraphics[angle=0, width=.4\textwidth]{BLOB_XeSn32_Irho.eps}\label{fig_phase_trans_b}}
}
\caption{
(a)
	Evolution of the size of potential ripples in $^{136}$Xe$+^{124}$Sn at 25 and 32$A$MeV in central collisions (adapted from ref.~\cite{Napolitani2013}).
Spinodal-like fragmentation occurring at around 80--100fm/c is followed by a process of recombination at later times.
(b)
	Average isovector density in potential ripples as a function of the local density averaged on potential ripples at different intervals of time. Different sets of potential-ripple sizes are selected and indicated by the corresponding mass $A$.
}\label{fig_phase_trans}
\end{figure}

Due to compression and/or thermal effects,
the composite systems formed in central heavy ion collisions at Fermi energies (30-60
$A$MeV)
may reach low density values, attaining the co-existence zone of the
nuclear matter phase diagram. 
In this situation, the system may
undergo a spontaneous phase separation, breaking up into several fragments \cite{rep,Colonna1994}, as a consequence of the development of mean-field spinodal instabilities. 
However, already in the high density phase, nucleon correlations 
are expected to be rather large, due to the huge amount of
two-body nucleon-nucleon collisions.  Hence some memory of these high density correlations
could be kept along the clusterisation process.
According to the theoretical description adopted here, clusters emerge essentially from the occurrence of mean-field instabilities. However, many-body correlations play an essential role in any case because they provide the seeds for the nucleon assembly into clusters.  The interplay between mean-field and correlations effects
can be investigated comparing the results obtained with SMF and with BLOB.
Indeed the latter has been conceived with the purpose of including fluctuations in full 
phase space, thus improving the treatment of fluctuations and correlations, 
but preserving, at the same time, mean-field features such as the dispersion relation for unstable modes. 
Fluctuations act on both isoscalar and isovector degrees of freedom.
It is interesting to underline that, in neutron-rich systems, the cluster formation mechanism also keeps the fingerprints of the isovector channel of the
nuclear effective interaction, which is related to the symmetry energy term
of the nuclear EoS \cite{epja,bao-an}.  Indeed one observes that the clusters (liquid drops) 
which emerge from the low-density nuclear matter have a lower $N/Z$ ratio, 
with respect to the surrounding nucleons and light particles.
This effect, the so-called isospin distillation, is connected to the density derivative of the symmetry energy and leads to the minimisation of the system potential energy \cite{baranPR}.

The mechanism of cluster formation by mean-field instabilities is 
explored in fig.~\ref{fig_phase_trans_a}, in the system $^{136}$Xe$+^{124}$Sn for central collisions at 25 and 32 $A$MeV; a probability map shows how ripples in the potential landscape evolve in size as a function of time [from ref.\cite{Napolitani2013}].
At around 100fm/c large sizes, corresponding to the whole composite system, coexists with small sizes (especially at 25 $A$MeV), which are consistent with the leading wavelength of the dispersion relation, i.e. about the size of nuclei in the region of oxygen and neon~\cite{Colonna1994}.
If in this latter situation all potential concavities could come apart into fragments, a pure signature of the spinodal decomposition would stand out as a set of equal-size fragments~\cite{Borderie2001}.
However, when the radial expansion of the system is not sufficient to outweigh the mean-field resilience, the density landscape continues to evolve towards a more compact shape so that, if the system still succeeds in disassembling, potential concavities can merge together, giving rise to larger fragments and producing asymmetric fragment configurations.
This process of recombination prevails at 25$A$MeV and weakens at larger incident energies.
Fig.~\ref{fig_phase_trans_b} shows the action of the isovector terms on the same system in a process of isospin distillation. 
An asystiff EOS is used in fig.~\ref{fig_phase_trans_a} and fig.~\ref{fig_phase_trans_b}. 
As a function of the local density averaged on potential ripples, and for different sizes of those latter, the average isospin distribution for the corresponding sites is studied at different intervals of time.
Different sets of potential-concavity sizes are indicated by extracting a corresponding mass $A$.
From an initial situation where the system is close to saturation density and the average isospin is defined by the target and projectile nuclei, densities drop to smaller values and the isospin distribution extends over a large range.
In particular, the smaller is the local density, the larger is the isospin measured in corresponding sites, as neutrons favour the most volatile phase.
Such ordering has been intensively studied within the SMF approach~\cite{baranPR}. 

%
%
%
\begin{figure}[b!]
\centerline{
	\includegraphics[angle=0, width=.65\textwidth]{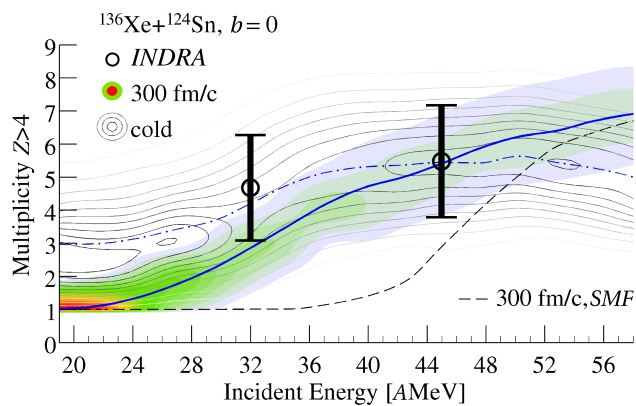}}
\caption{
	BLOB simulation: evolution of IMF ($Z>4$)  multiplicity, as a function of incident energy for the system $^{136}$Xe$+^{124}$Sn and a selection of central collisions at 300fm/c (colour shades) and for the cold system (grey contours). Corresponding mean values are indicated for the BLOB simulation and for a corresponding SMF calculation (adapted from ref.~\cite{Napolitani2013}).
	Corresponding experimental data from Indra experiments~\cite{Moisan2012,Ademard2014} are added for comparison, giving average (symbols) and variance of a gaussian fitted to the multiplicity distributions (bars).
}\label{fig_frag_threshold}
\end{figure}

%
%
%
\begin{figure}[t!]
\centerline{
	\includegraphics[angle=0, width=.7\textwidth]{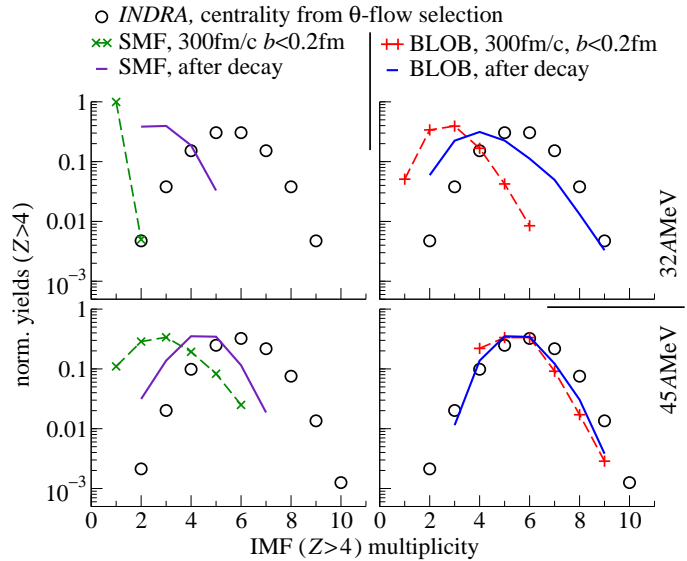}}
\caption{
	BLOB and SMF simulations: IMF ($Z>4$) multiplicity distributions for the systems $^{136}$Xe$+^{124}$Sn at 32 and 45$A$MeV, for a selection of central collisions at 300fm/c and after secondary decay
are compared to experimental data from Indra, with average multiplicity and variance given in Refs.~\cite{Moisan2012,Ademard2014}. 
The experimental distributions correspond to the bars shown in fig.~\ref{fig_frag_threshold}. 
}\label{fig_M_spectra}
\end{figure}

The improvement introduced with the BLOB approach is primarily providing a correct sampling of the isoscalar amplitudes and yielding a consistent description of the threshold toward multifragmentation.
Fig.~\ref{fig_frag_threshold} shows how in the system $^{136}$Xe$+^{124}$Sn, analysed for central impact parameters at 300fm/c, fragmentation events start competing with the predominant fusion mechanism already beyond 20 $A$MeV per nucleon of incident energy and, while the multiplicity of intermediate mass fragments 
(IMF) with $Z>4$, registered at 300 fm/c, continues to grow, it exhibits a maximum at around 45 $A$MeV when considering cold fragments.
A calculation undertaken with SMF for the same system results not well adapted when exploring too low incident energies, due to the smaller amplitude of the isoscalar fluctuations.
In the figure, two experimental points from INDRA~\cite{Moisan2012,Ademard2014} indicate the contribution from a compact source; they are compared to the calculated cold distribution and indicate that the BLOB simulation is quantitatively consistent.
The cooling of the hot system (300fm/c) is undertaken by the use of the decay model Simon~\cite{Durand1992}. 
Additional details on the comparison between theory and 
experiment~\cite{Moisan2012,Ademard2014} are provided in fig.~\ref{fig_M_spectra}, where the difference between SMF and BLOB is tested around the multifragmentation threshold (32$A$MeV) and in full multifragmentation regime (45$A$MeV). 

\section{Survey of results for semi-peripheral collisions}
%
%
%
\begin{figure}[b!]
\centerline{
	\includegraphics[angle=0, width=.75\textwidth]{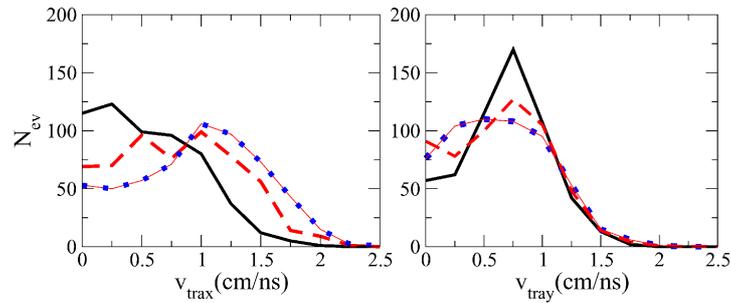}}
\caption{
Left panel: transverse velocity in reaction plane distribution for
fragmentation events with three IMFs, as obtained in the reaction
 $^{124}$Sn + $^{124}$Sn, at 50 $A$MeV, b = 4 fm. 
Right panel: transverse
velocity out of reaction plane distribution for the same set of events.
 Heaviest (black, continuous line), second heaviest (red,
long-dashed line), and third heaviest (blue,dotted line) fragments in the
hierarchy.
The asysoft EoS is adopted in the calculations. 
Adapted from Ref.\cite{Baran2012}. }
\label{transversethree}
\end{figure}
%
%
%
\begin{figure}[b!]
\centerline{
	\includegraphics[angle=0, width=.75\textwidth]{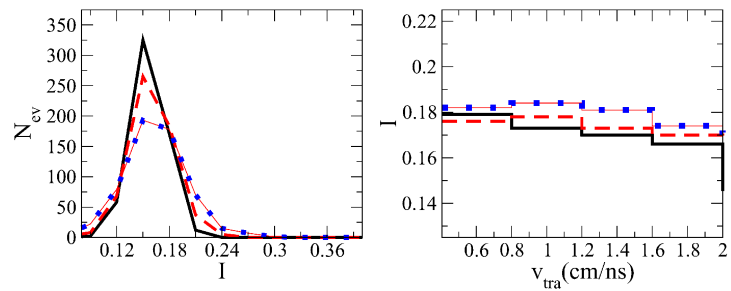}}
\caption{ Asysoft EOS
choice. Events with three IMFs. Left panel: Distribution of the
 isospin parameter $I = (N-Z)/A$.
Right panel: Fragment isospin content as a function of transverse
velocity. All lines are like in the caption of fig.~\ref{transversethree}.
Adapted from Ref.\cite{Baran2012}.}
\label{Asysoftthree}
\end{figure}
%
%
%
\begin{figure}[b!]
\centerline{
	\includegraphics[angle=0, width=.75\textwidth]{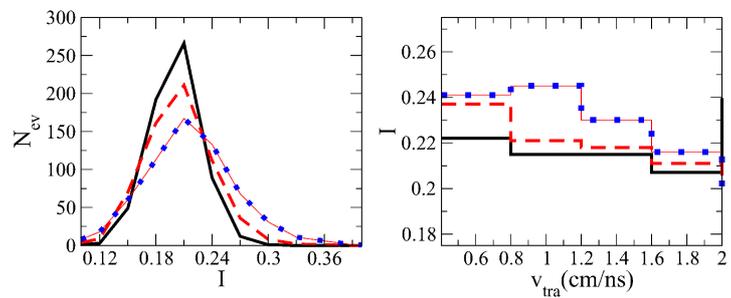}}
\caption{Like in fig.~\ref{Asysoftthree}
, for the asysuperstiff EOS. Adapted from Ref.\cite{Baran2012}.}
\label{Asysuperstiffthree}
\end{figure}

%

As discussed above, for central collisions, the nuclear multifragmentation can
be associated with a liquid-gas phase transition in a composite
system.
The kinetics of this phase transition is related to spinodal
decomposition in two-component nuclear matter 
accompanied by the isospin distillation. 
At semicentral impact parameters the mechanism changes from fusion to binary channels at low incident energy; in this transition an intermediate mechanism may appear where a neck is produced~\cite{Baran2012,Rizzo2014}.
The neck fragmentation with a peculiar intermediate
mass fragment ($2<Z<20$) distribution and an entrance	
channel memory was observed experimentally and
predicted by various transport models \cite{baranPR,Lionti2005,DiToro2006,Udo}. 
In this case,
the low-density neck region triggers an isospin migration from
the higher density regions corresponding to the projectilelike
fragment (PLF) and targetlike fragment (TLF). Therefore,
the isospin content of the IMFs is expected to reflect the
isospin enrichment of the midvelocity region. Many investigations have been carried out with the SMF approach~\cite{baranPR,Rizzo2008_1}. 
For even more
peripheral collisions, an essentially binary reaction in the
exit channel can by accompanied by a dynamically induced
fission of the participants, and for
N/Z-asymmetric
entrance channel combinations, isospin diffusion drives the
system toward charge equilibration \cite{betty,Rizzo2008_1}.
Consequently, the isospin degree of freedom can be seen
as a precious tracer providing additional information about
the physical processes taking place during the evolution
of the colliding systems. Moreover, from a comparison
between the experimental data and the theoretical model
predictions, isospin dynamics allows one to investigate the
density and/or temperature dependence of the symmetry
energy \cite{epja}. More exclusive analyses from the new experimental
facilities will certainly impose severe restrictions on various
models and parameterisations concerning this quantity.
An interplay between statistical and
dynamical mechanisms is expected and we clearly evidence
the development of hierarchy effects in the transverse velocity
of IMFs, which can be a signal of the cluster formation time scale
and of the importance of many-body correlations.  
Moreover, new interesting correlations between
kinematic features of the fragments and isospin dynamics,
which can provide clues in searching for the most sensitive
observables to the symmetry energy, are noted.

These effects are illustrated below, in the context of SMF and BLOB
models, for symmetric reactions. 

Fig.~\ref{transversethree} shows results obtained for the system $^{124}$Sn + $^{124}$Sn, at 50 $A$MeV, b = 4 fm, with the asysoft EoS. Transverse velocity distributions are presented for events with three IMF's, ordering the fragments, in each event, according to their size. One can see that the smallest fragment may acquire a large transverse velocity, especially on the reaction plane, indicating that it is emitted on shorter time scales.

The emission time has important consequences on the fragment features and, in
particular, on the isospin content.  Indeed the isospin migration mechanism
will be more effective for fragments which stay longer in contact with PLF
and TLF, as illustrated in figs~\ref{Asysoftthree} and ~\ref{Asysuperstiffthree}. 
We observe in fact that smaller fragments are generally more neutron-rich.
In addition, the isospin content decreases with the transverse velocity.
These effects are more pronounced in the 
asystiff case, which leads to the largest migration effects \cite{baranPR}. 

This is a clear example of the impact of the reaction dynamics and of the
cluster emission time scale, which is strongly influenced by correlation
effects, on the cluster properties.  
    

%
%
%
\begin{figure}[b!]
\centerline{
  \minifigure[
Bottom. Survey of average fragment multiplicity in peripheral collisions for BLOB and SMF simulations for the system $^{136}$Xe$+^{124}$Sn at 25, 32 and 45$A$MeV at 300fm/c.
The 32$A$MeV system is also shown after secondary decay.
Top. corresponding mass distributions at 32$A$MeV for the hot system.
]
	{\includegraphics[angle=0, width=.52\textwidth]{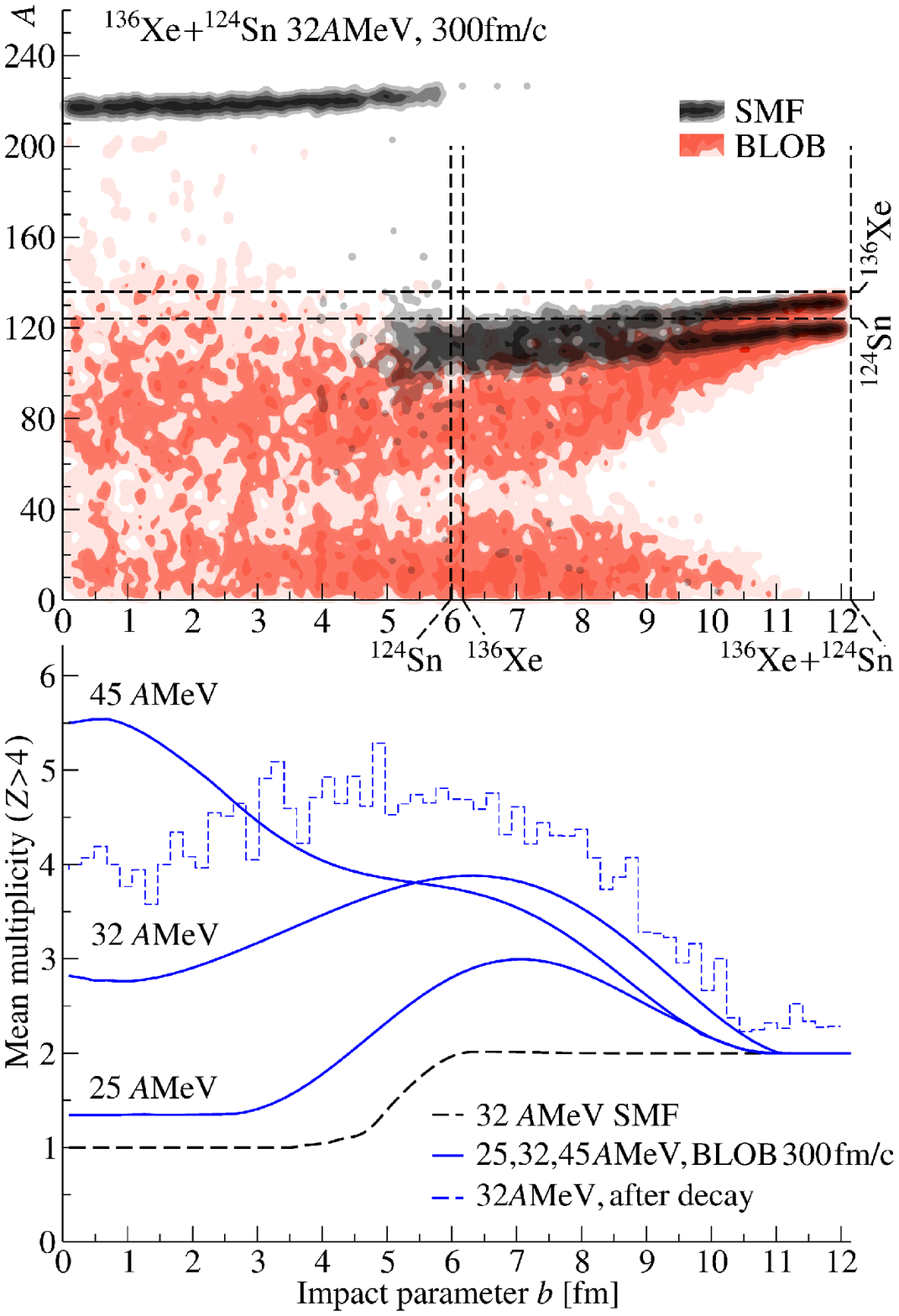}\label{fig_Neck_NoverZ_a}}
  \minifigure[
Isotopic content of neck fragments at 300fm/c, identified in the 
impact parameter region of maximum production, 
as a function of the transverse velocity component with respect to the quasi-projectile--quasi-target axis for a stiff and a soft form of the symmetry energy potential component. 
]
	{\includegraphics[angle=0, width=.4\textwidth]{NovZ_vper_XeSN32b6.eps}\label{fig_Neck_NoverZ_b}}
}
\end{figure}

Fig.~\ref{fig_Neck_NoverZ_a} extends the survey of fig.~\ref{fig_frag_threshold} to all impact parameters for the same system $^{136}$Xe$+^{124}$Sn.
The improvement that the BLOB approach introduces is not directly in the description of the isovector modes, but rather in the more complete description of the isoscalar modes.
Those latter allow for extending the mechanism of neck fragmentation down to low incident energies (below 15$A$MeV) and accessing isospin migration when the process of ternary splits is the most relevant for semiperipheral impact parameters.
This is illustrated in the bottom panel of fig.~\ref{fig_Neck_NoverZ_a}, where ternary and quaternary splits (related to neck formation) at 300fm/c are shown to dominate the semiperipheral impact parameters below 32$A$MeV.
A maximum of IMF multiplicity (quaternary splits) is reached at 32$A$MeV; on the contrary, at this same energy, the rate of ternary breakups in SMF is still negligible.

The top panel of fig.~\ref{fig_Neck_NoverZ_a}, adds information on the mass of the IMFs registered at 300fm/c for the 32$A$MeV incident energy.
While SMF jumps from fusion to binary splits, BLOB produces IMF at all impact parameters $b$, reaching a maximum of production above $b=6$fm for fragments with $A<40$.
This region is further studied in fig.~\ref{fig_Neck_NoverZ_b}, where the isotopic content $N/Z$ of the IMFs is shown as a function of the transverse velocity component with respect to the quasi-projectile--quasi-target axis for a stiff and a soft form of the symmetry-energy potential component.
The dramatically reduced production of IMF in SMF is accompanied by a much smaller mean transverse 
velocity component; this points to 
a more explosive dynamics in BLOB, resulting from the improved treatment of many-body correlations. 
The isovector features are expected from the properties of the employed interaction.
A stiff form of the symmetry-energy potential component is associated with 
a larger variation with density, thus 
enhancing the isospin migration towards the diluted neck region and 
resulting into a larger isospin content of the IMF. 
Smaller transverse velocity components are related to fragment configurations more aligned along the quasi-projectile--quasi-target axis, exploiting a longer interval of time available for the process of isospin migration, and finally favour more neutron rich fragments. 
The neutron enrichment reduces for less aligned configurations and larger transverse velocities.
This study is consistent with the experimental investigation reported in ref.~\cite{DeFilippo2005}.

\section{Spallation reactions induced by relativistic protons and deuterons}
%
%
%
\begin{figure}[b!]
\centerline{
	\includegraphics[angle=0, width=.65\textwidth]{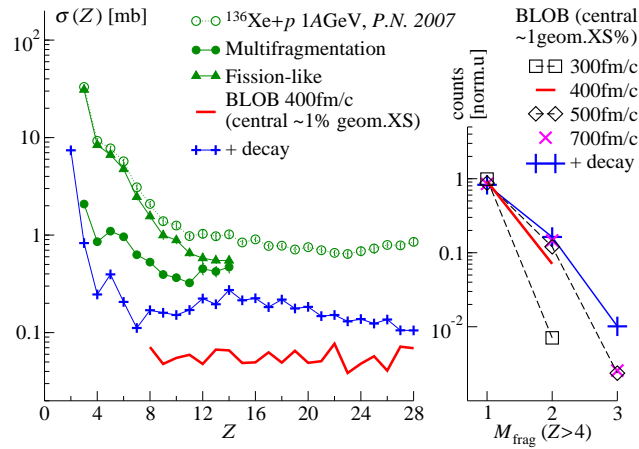}}
\caption{Left panel:
Production cross section of light elements emitted in $p+^{136}$Xe at 1$A$GeV.
Data were measured at the FRS (Darmstadt)~\cite{Napolitani2007}; contributions related to multifragmentation or to fission-like events could be separated~\cite{Napolitani2011}.
The full production of light elements is compared to a BLOB simulation restricted to central impact parameters covering 1\% of the total geometric cross section.
Right panel: corresponding multiplicity distribution, as obtained in BLOB calculations at different times and for the cold system. 
}\label{fig_spall_production}
\end{figure}


	In nuclear spallation reactions induced by relativistic light projectiles,
out-of-equilibrium processes are mainly associated to the very initial instants of the collision, 
when the light projectile traversing the heavy target induces a process of cavitation and the prompt emission of light ejectiles~\cite{Cugnon1981,Cugnon1987,Cugnon1997}.
	Since this initial stage does not produce mayor dynamical deformations, the successive evolution of the system may be efficiently assimilated to the decay of a fully thermalised source~\cite{ISIS2006}.
	Even though this two-stage picture yields a successful description of both the heavy nuclide and the light particle spectra, it is not fully compatible with the production of clusters and fragments of intermediate mass (around oxygen, neon and above).
	Those latter 
indeed
manifest a kinematics which is more consistent with multifragmentation than to asymmetric fission~\cite{Napolitani2011}.
	Statistical multifragmentation approaches could handle these features~\cite{Napolitani2004,Souza2009} in a macroscopic description, but a microscopic time-dependent picture of the process requires a transport formalisation where a source of dynamical fluctuations is included.
So, also in this case, clusters may emerge from the interplay between low-density instabilities, which can be encountered because of the thermal expansion,  and dynamical correlations. 
	For this purpose, stochastic transport approaches have been applied to heated systems where the initial conditions were prepared as the outcome of a schematic intranuclear cascade treatment~\cite{Colonna1997,Napolitani2015}.
	In this framework, it was found that after the light projectile leaves the target, this latter
undergoes a dynamical process which may progress for a rather long time.
	If the target nucleus is sufficiently excited and brought to rather low density, mechanical instabilities can set in and even amplify, 
like for the scenario of heavy-ion collisions at Fermi energies, discussed in the previous sections.
	Some of those inhomogeneities seldom prevail, separate into fragments, and leave the target nucleus.
	Alternatively, they may combine in larger blobs of matter so that, even when the system undergoes a multifragmentation event, the multiplicity of fragments is small or even reduced to two; this last situation is particularly interesting because it gives rise to a binary split from a process which is fundamentally different from ordinary asymmetric fission.
	Fig.~\ref{fig_spall_production} illustrates the light elements produced in $p+^{136}$Xe at 1$A$GeV, measured at the FRS (Darmstadt)~\cite{Napolitani2007} and indicates an experimental selection of contributions which are consistent either to multifragmentation events or to binary splits~\cite{Napolitani2011}; the data are compared qualitatively to a BLOB calculation restricted to central impact parameters covering 1\% of the total geometric cross section (left panel). It is also shown (right panel) that the calculated multiplicity of fragments is predominantly reduced to two and, more rarely, to three fragments.

%
%
%
\begin{figure}[b!]
\centerline{
	\includegraphics[angle=0, width=.95\textwidth]{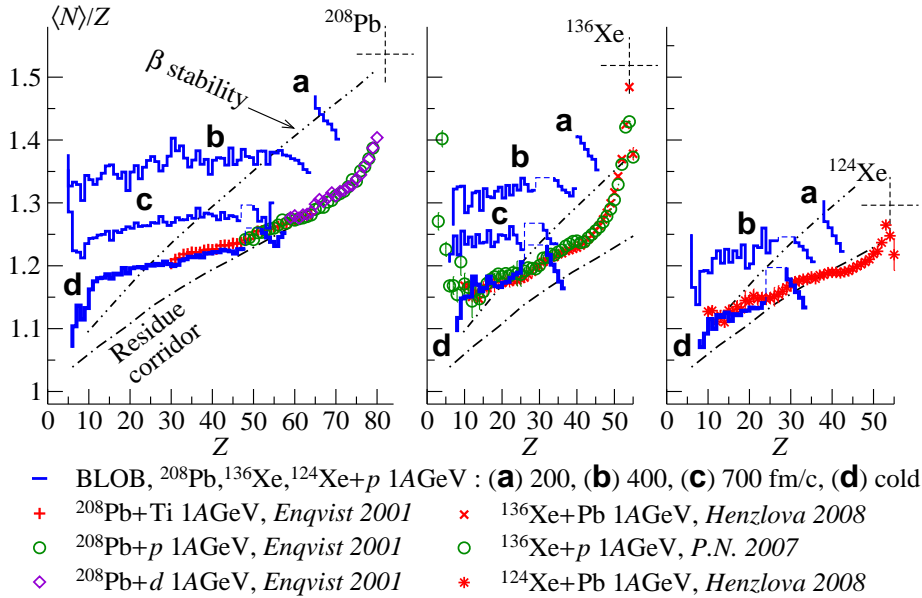}}
\caption{
Study of the evolution in time of the isotopic content of fragments produced in three spallation systems at relativistic energy.
The simulation, relying on the BLOB model~\cite{Napolitani2015}, is compared to several experimental data measured at FRS (Darmstadt) for fragments produced in proton/deuteron-induced spallation and for fragments emitted by spectator sectors in relativistic peripheral ion-ion collisions~\cite{Enqvist01,EnqvistTI01,Enqvist02,Napolitani2007,Henzlova2008}.
}\label{fig_spall_NoverZ}
\end{figure}


	This scenario of frustrated multifragmentation does not only recall heavy-ion collisions approaching Fermi energies, but it can also extend to the spectator region of peripheral heavy-ion collisions at relativistic energies.

	A study where several experimental data measured at FRS (Darmstadt) both from spallation reactions and from peripheral heavy-ion collisions are compared to a transport simulation relying on the BLOB model is presented in fig.~\ref{fig_spall_NoverZ} for the isotopic component of the nuclide production in 
 $p+^{124}$Xe, $^{136}$Xe, $^{208}$Pb
at 1$A$GeV~\cite{Enqvist01,EnqvistTI01,Enqvist02,Napolitani2007,Henzlova2008}.
	In particular, it is shown that the isotopic content of the lightest IMFs evolves in time from approximately the value of the target to a value which approaches but does not reach the residue corridor~\cite{Charity98}. 
This effect~\cite{Schmidt2002} and the collapsing of all datasets on the same ridge for a given isotopic content is a well known signature of the presence of a limiting temperature in the disassembling of the system, in connection with the liquid-gas phase transition of nuclear matter.
	This feature also demonstrates the compatibility of the transport approach, which does not impose any prior equilibrium condition, with a statistical picture of a thermalised system~\cite{Raduta2006}.

\section{Conclusions}

We have discussed a mechanism for cluster production based on the development of 
mean-field instabilities occurring for nuclear matter at low density
and moderate temperature.  These conditions can be reached during the expansion
phase of a heavy ion reaction at Fermi energy or due to thermal expansion 
of an excited nuclear source. 
Though the mechanism is driven by the mean-field, many-body correlations play
an essential role because they provide the initial seeds of the cluster
assembly. The relative importance of mean-field vs. correlations effects in spinodal
decomposition has been investigated employing and comparing the results of two
transport models: SMF and BLOB.  The latter implements fluctuations in full
phase space, thus improving the description of many-body correlations. 
As a result, one observes an increased clustering probability, shifting
the threshold for multifragmentation towards lower excitation energies, 
in agreement with experimental data. 
The clusters also acquire a large collective velocity.  Fragmentation features
are also explored for semi-peripheral reactions where new interesting aspects
emerge from the neck dynamics.  Also in this case events with larger IMF
multiplicity are obtained with BLOB, with also larger transverse velocity. 
Comparison with experimental data is in progress. 
Finally the BLOB model well describes also the possible occurrence of
clustering, counterbalanced by mean-field resilience effects, in hot
(but not initially diluted) sources, such as the nuclear systems formed in spallation reactions.  From these results it clearly emerges the important role
of many-body correlations even in describing processes which are driven by
mean-field instabilities. 
Further extensions of the model, towards a better description of the production
of very light clusters,  are envisaged.

\section{Acknowledgments}

This work for V. Baran was supported by a grant of the Romanian National
Authority for Scientific Research, CNCS - UEFISCDI, project number PN-II-ID-PCE-2011-3-0972.


\vspace{1cm}
\bibliographystyle{ws-rv-van}
\bibliography{ws-rv-sample}

\begin{thebibliography}{9}

%
\bibitem{Neg}
	J. W. Negele and D. Vautherin, 
	Phys. Rev. C \bbsty{5}{1472}{1972}.
%
\bibitem{Dec}
	J. Decharg\'e and D. Gogny, 
	Phys. Rev. C \bbsty{21}{1568}{1980}.
%
\bibitem{Rein}
	J.R. Stone and P.-G. Reinhard, 
	Prog. Part. Nucl. Phys. \bbsty{58}{587}{2007}, and refs. therein. 
%
\bibitem{Bend}
	M. Bender, P.-H. Heenen, and P.-G. Reinhard, 
	Rev. Mod. Phys. \bbsty{75}{121}{2003}, and refs. therein. 
%
\bibitem{Ebran2014}
	J.-P. Ebran, E. Khan, T. Nik\v{s}i\'{c} and D. Vretenar
	Phys. Rev. C \bbsty{90}{054329}{2014}.

\bibitem{Kanada2012} 
	Y. Kanada-En'yo, Masaaki Kimura and Akira Ono
	Prog. Theor. Exp. Phys. \bbsty{}{01A202}{2012}.
%
\bibitem{Lattimer2000}
	J.M. Lattimer and M. Prakash, 
	Phys. Rep. \bbsty{333}{121}{2000}.
%
\bibitem{Frankland}
	R.Botet et al., 
	Phys. Rev. Lett. \bbsty{86}{3514}{2001}.
%
\bibitem{Moretto} J.B.Elliott et al.,  
	Phys. Rev. Lett. \bbsty{88}{042701}{2002}. 
%
	\bibitem{Dagostino} 
	M.D'Agostino et al., Nucl. Phys. \bbsty{A699}{795}{2002}. 
%
\bibitem{EPJA}
	G.Tabacaru et al., 
	Eur. Phys. J. A \bbsty{18}{103}{2003}.
%
\bibitem{rep} 
	Ph.Chomaz, M.Colonna and J.Randrup, 
	Phys. Rep. \bbsty{389}{263}{2004}.
%

%
\bibitem{Ono}
	A.Ono, 
	Phys. Rev. C \bbsty{59}{1999}{853}.
%
\bibitem{Bertsch_rep}
	G.F. Bertsch and S. Das Gupta, 
	Phys. Rep. \bbsty{160}{189}{1988}.
%
\bibitem{Aldo_rep}
	A.Bonasera, F.Gulminelli and J.Molitoris, 
	Phys. Rep. \bbsty{243}{1}{1994}.
%
\bibitem{md}
	J.Aichelin, 
	Phys. Rep. \bbsty{202}{1991}{233}.
%
\bibitem{feld} 
	H.Feldmeier, 
	Nucl.Phys. \bbsty{A515}{147}{1990}.
%
\bibitem{ONOab}
	A. Ono, H. Horiuchi, T. Maruyama and A. Ohnishi,
	Phys. Rev. Lett. \bbsty{68}{2898}{1992};
	A. Ono, H. Horiuchi, T. Maruyama and A. Ohnishi,
	Prog. Theor. Phys. \bbsty{87}{1185}{1992}.
%
\bibitem{Pap} 
	M.Papa, T.Maruyama and A.Bonasera,
	Phys. Rev. C \bbsty{C64}{024612}{2001}.
%
\bibitem{FFMD}
	M.Colonna and Ph.Chomaz, 
	Phys. Lett. B \bbsty{436}{1}{1998}.
%
\bibitem{Ayik} 
	S. Ayik, C. Gregoire, 
	Phys. Lett. B \bbsty{212}{269}{1988};
	S.Ayik and C.Gregoire, 
	Nucl. Phys. \bbsty{A513}{187}{1990}.
%
\bibitem{Ayik1} 
	Y. Abe, S. Ayik, P.-G. Reinhard and E. Suraud, 
	Phys. Rep. \bbsty{275}{49}{1996}.
%
\bibitem{Randrup} 
	J.Randrup and B.Remaud, 
	Nucl. Phys. \bbsty{A514}{339}{1990}. 
%
\bibitem{Akira_comparison}
	J.Rizzo, M.Colonna, A.Ono, 
	Phys. Rev. C \bbsty{76}{024611}{2007}.
%
\bibitem{Salvo}
	M.Colonna {et al.},  
	Nucl. Phys. \bbsty{A642}{449}{1998}.
%
\bibitem{Napolitani2013}
	P. Napolitani and M. Colonna,
	Phys. Lett. B \bbsty{726}{382}{2013}.
%
\bibitem{Bauer1987}
	W. Bauer, G.F. Bertsch and S. Das Gupta,
	Phys. Rev. Lett. \bbsty{58}{863}{1987}.
%
\bibitem{Chapelle1992}
	F. Chapelle, G.F. Burgio, Ph. Chomaz and J. Randrup,
		Nucl. Phys. \bbsty{A540}{227}{1992}.
%
\bibitem{Rizzo2008}
	J. Rizzo, Ph. Chomaz and M. Colonna,
	Nucl. Phys. \bbsty{A806}{40}{2008}.
%
\bibitem{Napolitani_IWM2014}
	P. Napolitani, M. Colonna, V. de la Mota,
	Conf. proc. IWM2014-EC, Catania 2014,
	EPJ Web of Conf. \bbsty{88}{00003}{2015}.
%
\bibitem{Colonna1994}
	M. Colonna and Ph. Chomaz,
	Phys. Rev. C \bbsty{49}{1908}{1994}.
%
\bibitem{Napolitani2012}
	P. Napolitani and M. Colonna,
	Conf. proc. IWM2011-EC, Caen 2011,
	EPJ Web of Conf. \bbsty{31}{00027}{2012}.
%
\bibitem{Borderie}
	B. Borderie, M.F. Rivet,
	Progr. in Part. and Nucl. Phys. \bbsty{61}{(Book Series), 551}{2008}, and refs. therein.
%
\bibitem{TWINGO}
	A. Guarnera, M. Colonna, Ph. Chomaz, 
	Phys. Lett. B \bbsty{373}{297}{1996}. 
%
\bibitem{epja}M.Colonna, 
	V. Baran and M. Di Toro, 
	Eur. Phys. J. A \bbsty{50}{No.2, 30}{2014}.
%
\bibitem{baranPR} 
	V. Baran, M. Colonna, V. Greco  and M. Di Toro,  
	Phys. Rep. \bbsty{410}{335}{2005}.  
%
\bibitem{Joseph}
	J. Rizzo, M. Colonna and M. Di Toro, 
	Phys. Rev.  C \bbsty{72}{064609}{2005}.
%
\bibitem{DanielewiczCoupland}
P.Danielewicz,
Acta Phys.Polon. \bbsty{B33}{45}{2002} 
%
\bibitem{bao-an}
B.A. Li, L.W. Chen and C.M. Ko,
Phys. Rep. \bbsty{465}{113}{2008}.
%
\bibitem{Borderie2001} 
	B. Borderie et al. (INDRA Collaboration),
	Phys. Rev. Lett. \bbsty{86}{3252}{2001}.

\bibitem{Moisan2012} 
	F. Gagnon-Moisan et al. (INDRA Collaboration),
	Phys. Rev. C \bbsty{86}{044617}{2012}.
%
\bibitem{Ademard2014}
	G. Ademard et al. (INDRA Collaboration),
	Eur. Phys. J. A \bbsty{50}{33}{2014}.
%

%
\bibitem{Durand1992}
	D. Durand, 
	Nucl. Phys. \bbsty{A541}{226}{1992}.
%
\bibitem{Baran2012}	
	V. Baran, M. Colonna, M. Di Toro, and R. Zus,
	Phys. Rev. C \bbsty{85}{054611}{2012}.
%
\bibitem{Rizzo2014}	
	C. Rizzo, M. Colonna, V. Baran, and M. Di Toro
	Phys. Rev. C \bbsty{90}{054618}{2014}.
%
\bibitem{Lionti2005} 
	R. Lionti, V. Baran, M. Colonna and M. Di Toro,
	Phys. Rev. C \bbsty{71}{044602}{2005}.
%
\bibitem{DiToro2006} 
	M. Di Toro, A. Olmi and R. Roy,
	Eur. Phys. J. A \bbsty{30}{65}{2006} and refs. therein.

\bibitem{Udo}
B. Djerroud  et al.,
Phys. Rev. C \bbsty{64}{034603}{2001}; 
 J. Colin et al. (INDRA Coll.), Phys. Rev. C
\bbsty{67}{064603}{2003};
 A.A. Stefanini et al. Z. Phys. A
\bbsty{351}{167}{1995}. 
%

%
%
\bibitem{Rizzo2008_1} 
	J. Rizzo, M. Colonna, V. Baran, M. Di Toro, H.H. Wolter, M. Zielinska-Pfabe
	Nucl. Phys. \bbsty{A806}{79}{2008}.

\bibitem{betty}
M.B. Tsang et al., Phys. Rev. Lett.
\bbsty{102}
{122701}{2009}.

%
\bibitem{DeFilippo2005} 
	De Filippo E. et al.,
	{Phys. Rev. C} \bbsty{71}{044602}{2005}.



%
\bibitem{Cugnon1981}
	J. Cugnon, T. Mizutani, J. Vandermeulen,
	Nucl. Phys. \bbsty{A352}{505}{1981}.
%
\bibitem{Cugnon1987}
	J. Cugnon, 
	Nucl. Phys. \bbsty{A462}{751}{1987}.
%
\bibitem{Cugnon1997}
	J. Cugnon, C. Volant, S. Vuillier,
	Nucl. Phys. \bbsty{A620}{475}{1997}.
%
\bibitem{ISIS2006}  
	V.E. Viola et al.,
	Phys. Rep. \bbsty{434}{1}{2006}.
%
\bibitem{Napolitani2011} 
	P.Napolitani, K.-H.Schmidt and L.Tassan-Got,
	J. Phys. G: Nucl. Part. Phys. \bbsty{38}{115006}{2011}.
%
\bibitem{Napolitani2004} 
	P. Napolitani, K.-H. Schmidt, A.S. Botvina, F. Rejmund, L. Tassan-Got and C.Villagrasa,
	Phys. Rev. C \bbsty{70}{054607}{2004}.
%
\bibitem{Souza2009} 
	S.R. Souza, B.V. Carlson, R. Donangelo, W.G. Lynch, A.W. Steiner and M.B. Tsang,
	Phys. Rev. C \bbsty{79}{054602}{2009}.
%
\bibitem{Colonna1997} 
	M. Colonna, J. Cugnon, and E.C. Pollacco,
	Phys. Rev. C \bbsty{55}{1404}{1997}.
%
\bibitem{Napolitani2015} 
	P. Napolitani and M. Colonna,
	Phys. Rev. C \bbsty{92}{034}{2015}.

%
\bibitem{Napolitani2007} 
	P. Napolitani et al.,
	Phys. Rev. C \bbsty{76}{064609}{2007}.


%
\bibitem{Enqvist01} 
	T. Enqvist et al.,
	Nucl. Phys. \bbsty{A686}{481}{2001}.
%
\bibitem{EnqvistTI01} 
	P. Napolitani et al.,
	Conf. proc. VII Int. School Seminar on Heavy-Ion Phyics, Dubna,
	Phys. Atom. Nucl. \bbsty{66}{1471}{2003}.
%
\bibitem{Enqvist02} 
	T. Enqvist et al.,
	Nucl. Phys. \bbsty{A703}{435}{2002}.
%
\bibitem{Henzlova2008} 
	D. Henzlova et al.,
	Phys. Rev. C \bbsty{78}{044616}{2008}. 
%
\bibitem{Charity98} 
	R.J. Charity,
	Phys. Rev. \bbsty{58}{1073}{1998}.
%
\bibitem{Schmidt2002} 
	K.-H. Schmidt, M.V. Ricciardi, A.S. Botvina, T. Enqvist,
	Nucl. Phys. \bbsty{A710}{157}{2002}.
%
\bibitem{Raduta2006} 
	A.H. Raduta, M. Colonna, V. Baran and M. Di Toro,
	Phys. Rev. C \bbsty{74}{034604}{2006}.
%

\end{thebibliography}

\end{document}